\documentclass[aps,pre,twocolumn,groupedaddress]{revtex4-1}
\usepackage{color}
\usepackage{tabularx}
\usepackage{epsfig}
\usepackage{amsmath}
\usepackage{graphicx}

\newcommand{\smean}[1]{\left\lbrack #1 \right\rbrack}
\newcommand{\mean}[1]{\left\langle #1 \right\rangle}
\newcommand{\var}[1]{\mathrm{var}\!\left( #1 \right)}

\begin{document}

\title{Non-self-averaging in Ising spin glasses; hyperuniversality}

\author{P. H.~Lundow} \affiliation{Department of Mathematics and
  Mathematical Statistics, Ume{\aa} University, SE-901 87, Sweden}

\author{I. A.~Campbell} \affiliation{Laboratoire Charles Coulomb
  (L2C), UMR 5221 CNRS-Universit\'e de Montpellier, Montpellier,
  F-France.}

\date{\today}

\begin{abstract}
  Ising spin glasses with bimodal and Gaussian near-neighbor
  interaction distributions are studied through numerical simulations.
  The non-self-averaging (normalized inter-sample variance) parameter
  $U_{22}(T,L)$ for the spin glass susceptibility (and for higher
  moments $U_{nn}(T,L)$) is reported for dimensions $2, 3, 4, 5$ and
  $7$. In each dimension $d$ the non-self-averaging parameters in the
  paramagnetic regime vary with the sample size $L$ and the
  correlation length $\xi(T,L)$ as $U_{nn}(\beta,L) =
  [K_{d}\xi(T,L)/L]^d$, and so follow a renormalization group law due
  to Aharony and Harris \cite{aharony:96}. Empirically, it is found
  that the $K_{d}$ values are independent of $d$ to within the
  statistics. The maximum values $[U_{nn}(T,L)]_{\max}$ are almost
  independent of $L$ in each dimension, and remarkably the estimated
  thermodynamic limit critical $[U_{nn}(T,L)]_{\max}$ peak values are
  also dimension-independent to within the statistics and so are
  "hyperuniversal".  These results show that the form of the spin-spin
  correlation function distribution at criticality in the large $L$
  limit is independent of dimension within the ISG family. Inspection
  of published non-self-averaging data for $3$D Heisenberg and XY spin
  glasses the light of the Ising spin glass non-self-averaging results
  show behavior incompatible with a spin-driven ordering scenario, but
  compatible with that expected on a chiral-driven ordering
  interpretation.
\end{abstract}


\maketitle

\section{Introduction}
The non-self-averaging parameter, usually noted $A$ or $U_{22}$,
represents the normalized inter-sample variability for systems such as
diluted ferromagnets or spin glasses where the microscopic structures
of the interactions within individual samples are not identical. The
parameter is defined for ferromagnets as the inter-sample variance of the
susceptibility normalized by the mean susceptibility squared
\cite{aharony:96},
\begin{equation}
  U_{22}(\beta,L) = \left(\frac{\sigma_{\chi}(T,L)}{\chi(T,L)}\right)^2 = 
  \frac{\var{\mean{q^2}}}{\smean{\mean{q^2}}^2}
  \label{U22def}
\end{equation}
where $\sigma(T,L)$ is the standard deviation of the equilibrium
sample-by-sample distribution of the susceptibility.  We denote by
$\mean{\cdots}$ the thermal mean for a single sample and by
$\smean{\cdots}$ the sample mean.  In ISGs the spin glass
susceptibility replaces $\chi$.  The non-self-averaging definition can
be widened to other observables \cite{aharony:96}; we will also
discuss the behavior of non-self-averging of higher moments
$\mean{q^3}$ and $\mean{q^4}$ of of the spin-spin correlation $q$.

Aharony and Harris \cite{aharony:96} gave a fundamental
renormalization group discussion of non-self-averaging in diluted
ferromagnets, which can be applied also to spin glass models. First,
they showed that in the paramagnetic regime, at temperatures above the
critical temperature, $U_{22}$ (which they referred to as $R_{\chi}$)
behaves as
\begin{equation}
U_{22}(T,L) \sim (\xi(T,L)/L)^d
\label{U22xiL}
\end{equation}
where $d$ is the dimension of the system. This rule can be understood
on a simple physical picture : the inter-sample variability depends on
the ratio of the sample volume to the correlated volume. Roughly, each
sample is contained in a "box" of volume $L^d$. When this box volume
is much larger than the correlated volume $\xi(T)^d$, all samples will
have essentially identical properties; when the inverse is true, each
sample has its own individual properties.

Then at the critical point $T_{c}$ where $\xi(T)$ diverges in the
thermodynamic limit ThL, $U_{22}(T_{c},L)$ becomes independent of $L$
even when $L$ tends to infinity \cite{aharony:96}. In this strongly
non-self-averaging regime the observables for each individual sample
have different properties. The passage as a function of temperature in
the thermodynamic limit from "all samples identical" (randomness
irrelevant) to "all samples different" (randomness relevant) is a
fundamental signature of the physical meaning of ordering in systems
with disorder or in spin-glass-like systems. Aharony and Harris show
that the value of $U_{22}(T_{c},L)$ in the limit of large $L$ should
be universal, for ferromagnets with different forms of disorder in a
given dimension. We find empirically that within the ISG family this
critical parameter is dimension-independent, i.e. hyperuniversal.

We report non-self-averaging measurements in near neighbor interaction
Ising spin glasses (ISGs) having dimensions $2, 3, 4, 5$ and $7$, with
bimodal or Gaussian near neighbor interaction distributions. There is
considerable regularity in behavior throughout all this range of $d$,
which includes the special cases $d=2$ where $T_{c}=0$, and $d=7$
which is above the upper critical dimension $d=6$. In the paramagnetic
regime $U_{22}(T,L) = [K_{d}\xi(T,L)/L]^d$ with $K_{d} \approx 2.5$
for all $d$ studied, to within the statistical accuracy. Secondly, the
peak in $U_{22}(T,L)$ as a function of $T$ for fixed $L$ has a value
$U_{22}(\max)$ for each $L$ which, after weak small size effects, is
independent of $L$ and also of $d$ to within the statistics,
$U_{22}(\max)\approx 0.205$. The location of the peak
$T(U_{22}(\max))$ approaches $T_{c}$ from the paramagnetic regime
(higher $T$) for $d < 4$ and from the ordered regime (lower $T$) for
$d > 4$. The same rules are followed for the higher moments of the
spin-spin correlations.

\section{Simulations}

The standard ISG Hamiltonian is
\begin{equation}
  \mathcal{H}= - \sum_{ij}J_{ij}S_{i}S_{j}
  \label{ham}
\end{equation}
with the near neighbor symmetric distributions normalized to $\langle
J_{ij}^2\rangle=1$. The normalized inverse temperature is $\beta =
(\langle J_{ij}^2\rangle/T^2)^{1/2}$. The Ising spins live on simple
hyper-cubic lattices with periodic boundary conditions. The spin
overlap parameter is defined as usual by
\begin{equation}
  q =\frac{1}{L^{d}}\sum_{i} S^{A}_{i} S^{B}_{i}
\end{equation}
where the sum is taken over all spins and A, B indicate two copies of
the same system.  The spin glass susceptibility is then defined as
usual $\chi(\beta,L) = L^d \lbrack\mean{q^2}\rbrack$.  

The equilibration techniques (which are different in dimension $2$)
are described in Refs.~\cite{lundow:15,lundow:15a}.  On a technical
level, it turns out that the values of $U_{22}$ and particularly the
peak value can fluctuate slightly in an irregular manner as at each
size they depend sensitively on strict equilibration having been
achieved. This can be used as a convenient empirical test for
equilibration.

\section{Dimension 2}

It is well established that short range ISGs in dimension 2 only order
at $T=0$ \cite{hartmann:01,ohzeki:09}.  The Gaussian ISG has a
non-degenerate ground state and a continuous energy level
distribution.  The bimodal ISG has an effectively continuous energy
level regime down to an $L$ dependent cross-over temperature
$T^{*}(L)$ below which the thermodynamics are dominated by the
massively degenerate ground state \cite{jorg:06}. This is a finite
size regime; in the thermodynamic limit regime the bimodal ISG can be
considered to have an effectively continuous energy level distribution
similar to that of the Gaussian ISG.

\begin{figure}
  \includegraphics[width=3.0in]{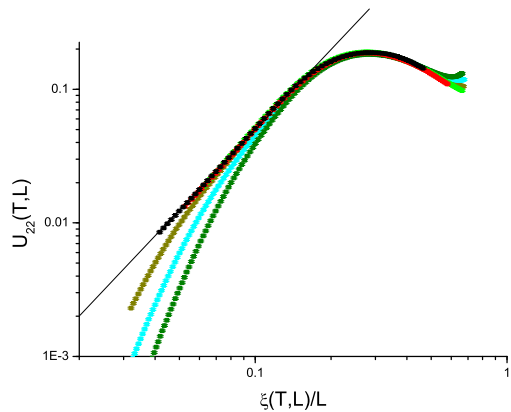}
  \caption{(Color on line) Bimodal $2$D ISG. Non-self-averaging
    parameter $U_{22}(T,L)$ against the normalized correlation length
    $\xi(T,L)/L$. $L = 12$, $16$, $24$, $32$, $48$, $64$, $96$, $128$
    from right to left. The straight line has slope $2$.}
  \protect\label{fig1}
\end{figure}

Measurements on two bimodal models and the Gaussian model ISG in
dimension 2 \cite{toldin:10} show a clear scaling of $U_{22}(T,L)$ as
a function of $\xi(T,L)/L$, with all the maxima in $U_{22}(T,L)$ close
to $0.20$. We show for the standard bimodal ISG in dimension 2,
Fig.~\ref{fig1}, the data scaled against $\xi(T,L)/L$ on a log-log
plot. This brings out the fact (not mentioned in
Ref.~\cite{toldin:10}) that for temperatures above the peak location
temperature, the Aharony-Harris rule \cite{aharony:96} $U_{22}(T,L) =
(K_{2}\xi(T,L)/L)^2$ holds, with $K_{2}= 2.5(1)$. Below the peak
obvious finite size effects due to the crossover to the ground state
dominated regime set in.

\begin{figure}
  \includegraphics[width=3.0in]{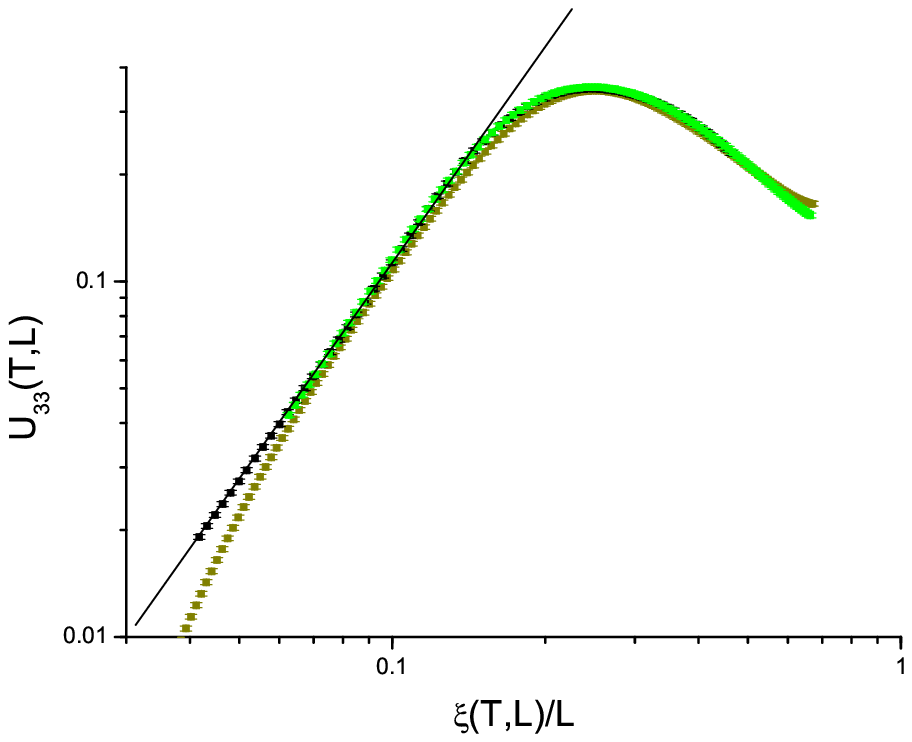}
  \caption{(Color on line) Bimodal $2$D ISG. Non-self-averaging
    parameter $U_{33}(T,L)$ against the normalized correlation length
    $\xi(T,L)/L$. $L = 24$, $32$, $48$, $64$, $96$, $128$ from right to
    left. The straight line has slope $2$.} \protect\label{fig2}
\end{figure}

\begin{figure}
  \includegraphics[width=3.0in]{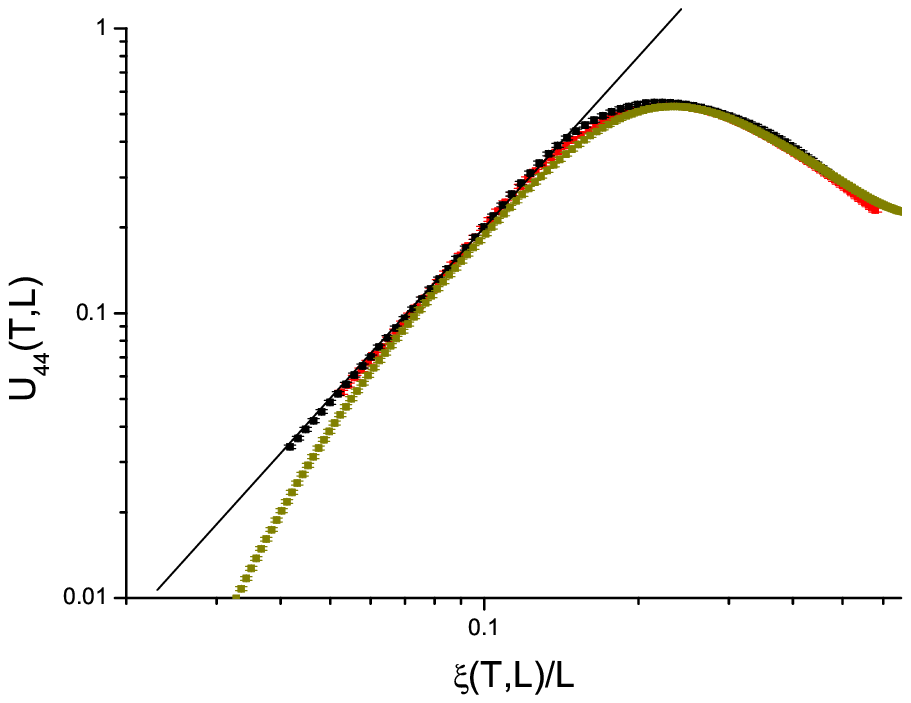}
  \caption{(Color on line)Bimodal $2$D ISG. Non-self-averaging
    parameter $U_{44}(T,L)$ against the normalized correlation length
    $\xi(T,L)/L$. $L = 24$, $48$, $64$, $128$ from right to
    left. The straight line has slope $2$.} \protect\label{fig3}
\end{figure}

From the same simulation runs, data for the higher moments
$\mean{q^3}(T,L)$ and $\mean{q^4}(T,L)$ were obtained and the values
of the normalized variances $U_{33}(T,L)$ and $U_{44}(T,L)$ were
evaluated. Equivalent plots to Fig.~\ref{fig1} are shown for
$U_{33}(T,L)$ and $U_{44}(T,L)$ in Figs.~\ref{fig2} and \ref{fig3}
with $U_{33}(T,L) = (3.29\xi(T,L)/L)^2$ and $U_{44}(T,L) =
(4.36\xi(T,L)/L)^2$.

\begin{figure}
  \includegraphics[width=3.0in]{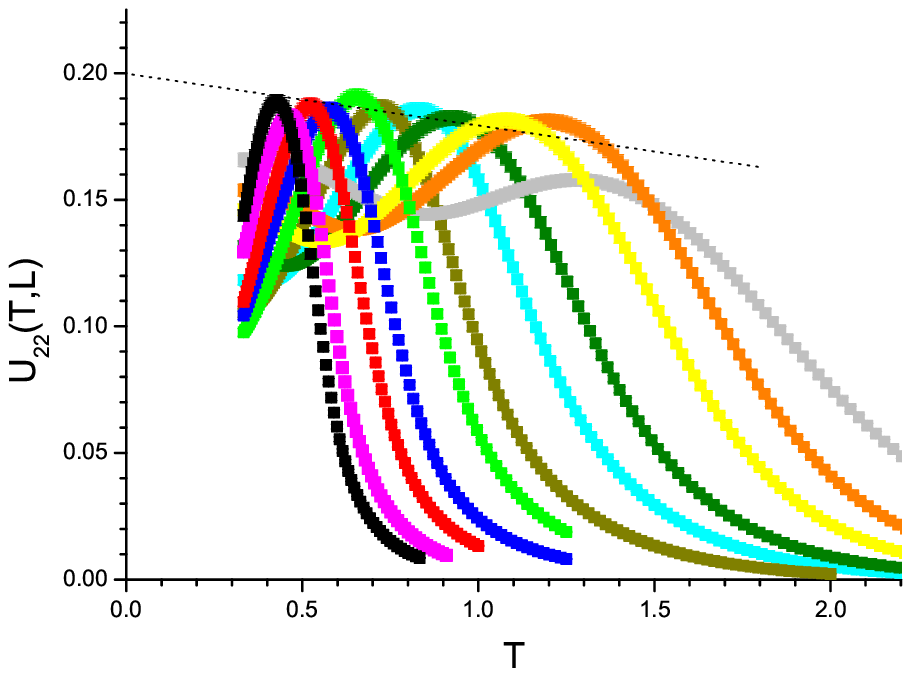}
  \caption{(Color on line) Bimodal $2$D ISG. Non-self-averaging
    parameter $U_{22}(T,L)$ against the temperature $T$. $L = 128$,
    $96$, $64$, $48$, $32$, $24$, $16$, $12$, $8$, $6$, $4$ from left
    to right. The straight line extrapolates to criticality at $T=0$.}
  \protect\label{fig4}
\end{figure}

The same data are presented as $U_{22}(T,L)$ against $T$ for fixed $L$
in Fig.~\ref{fig4}; the peak location is moving towards $T=0$ with increasing
$L$, and the maximum value is very gradually growing with increasing
$L$. A simple extrapolation of the peak data from $L=4$ to $L=128$
indicates a limiting infinite $L$ peak value close to $0.200$.

\begin{figure}
  \includegraphics[width=3.0in]{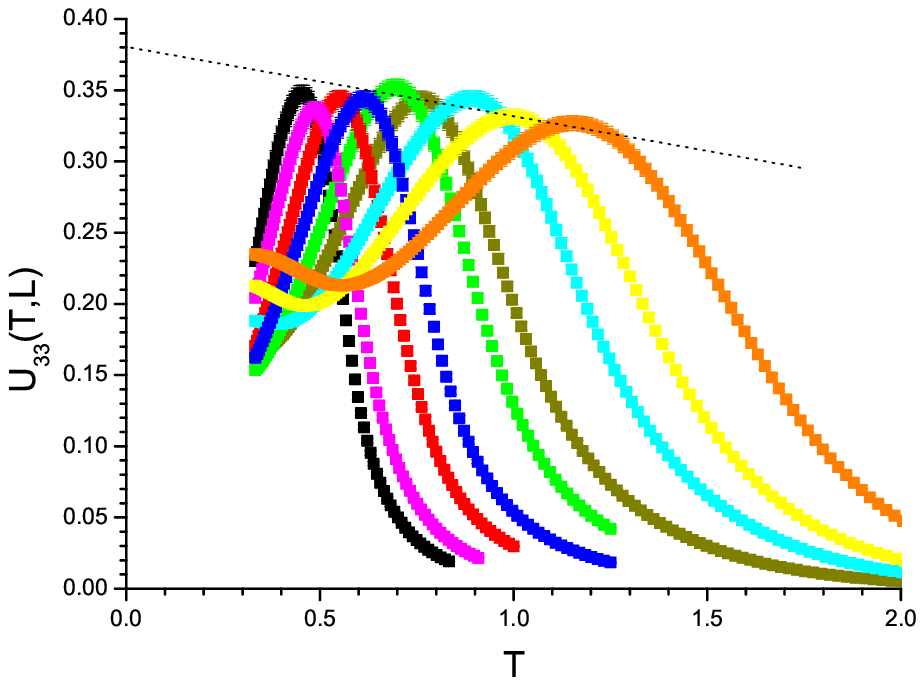}
  \caption{(Color on line) Bimodal $2$D ISG. Non-self-averaging
    parameter $U_{33}(T,L)$ against the temperature $T$. $L = 128$,
    $96$, $64$, $48$, $32$, $24$, $16$, $12$, $8$, $6$ from left to
    right. The straight line extrapolates to criticality at $T=0$.}
  \protect\label{fig5}
\end{figure}

\begin{figure}
  \includegraphics[width=3.0in]{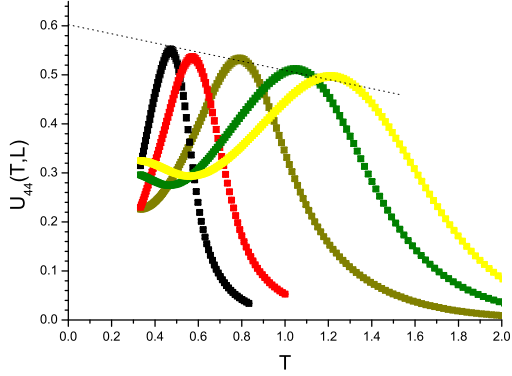}
  \caption{(Color on line) Bimodal $2$D ISG. Non-self-averaging
    parameter $U_{44}(T,L)$ against the temperature $T$. $L = 8$,
    $16$, $32$, $64$, $128$ from right to left. The straight line
    extrapolates to criticality at $T=0$.} \protect\label{fig6}
\end{figure}

The $U_{33}(T,L)$ and $U_{44}(T,L)$ peak values evolve in a very
similar way to the $U_{22}(T,L)$ peaks, extrapolating to large $L$
limit values $U_{33}=0.38(1)$ and $U_{44} = 0.60(1)$, Figs.~\ref{fig5}
and \ref{fig6}.  On the low temperature side of the bimodal data, a
minimum in each of the $U_{nn}(T,L)$ at an $L$ dependent temperature
followed by a plateau (see \cite{toldin:10}) provides a clear
indication of the crossover from the effectively continuous energy
level regime to the degenerate ground state dominated regime. For the
largest sizes, this crossover lies below the lowest temperatures at
which measurements were carried out.

Data for $U_{22}(T,L)$ for the Gaussian ISG (not shown) are very
similar to the bimodal data, except that there is of course no
crossover effect.

The zero temperature infinite size limit can be defined in two
ways. Taking the successive limits $L \to \infty, T \to 0$ gives an
extrapolated value $U_{22}(0,\infty) =0.200(5)$ for both bimodal and
Gaussian models, while the successive limits $T \to 0, L \to \infty$
gives a value $\approx 0$ in the Gaussian case; on the present data it
is hard to estimate in the bimodal model.

\section{Dimension 3}

The bimodal ISG in dimension 3 has a transition temperature for which
the most recent estimate is $T_{c}=1.102(3)$
\cite{katzgraber:06,hasenbusch:08,baity:13}, and the Gaussian ISG has
a transition temperature estimated to be $T_{c}=0.951(9)$
\cite{katzgraber:06}. The critical values of the dimensionless
correlation length ratio $[\xi(T,L)/L]_{c}$ are estimated to be
$0.652(3)$ and $0.635(10)$ respectively.

Hasenbusch, Pellisetto and Vicari \cite{hasenbusch:08} have generously
posted their raw tabulated simulation data for the bimodal ISG in
dimension 3 on the EPAPS site corresponding to their publication.  In
addition to the present measurements on $2^{13}$ samples of sizes $L=
4, 6, 8, 10, 12$ we have extracted a selection of values of
$U_{22}(\beta,L)$ from the tables of \cite{hasenbusch:08}, choosing
the data sets with the largest numbers of temperatures, $L=16$, $20$,
$24$. In each case the data correspond to measurements on about $10^5$
samples.

The $U_{22}(T,L)$ bimodal data in $3$D have almost $L$-independent
peak values $U_{22}(T,L)_{\max} =0.207(3)$ with peak locations tending
gradually downwards towards $T_{c}$ as $L$ increases, Fig.~\ref{fig8} (see
\cite{palassini:03} who observed also a very similar peak height for a
next-nearest-neighbor model). Small fluctuations as a function of $L$
can be put down to residual equilibration differences as the
statistical errors in these data are very small because of the large
numbers of samples. At the critical temperature the finite size
scaling limit is $U_{22}(T_{c},L) = 0.147$ \cite{hasenbusch:08}.

\begin{figure}
  \includegraphics[width=3.0in]{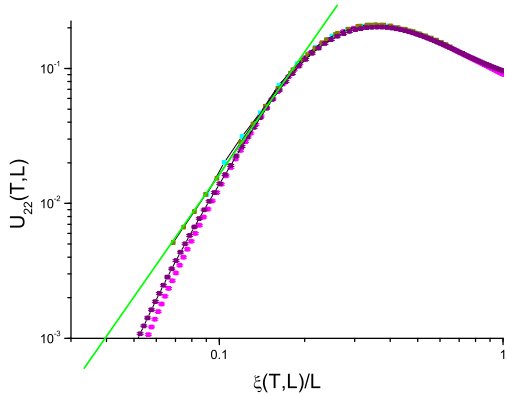}
  \caption{(Color on line) Bimodal 3D ISG. Non-self-averaging
    parameter $U_{22}(T,L)$ against the normalized correlation length
    $\xi(T,L)/L$. $L = 10$, $12$, $16$, $20$, $24$ (pink, purple, cyan,
    green, black) right to left.  $L = 16$, $20$, $24$ from
    \cite{hasenbusch:08}.  The straight line has slope 3.}
  \protect\label{fig7}
\end{figure}

\begin{figure}
  \includegraphics[width=3.0in]{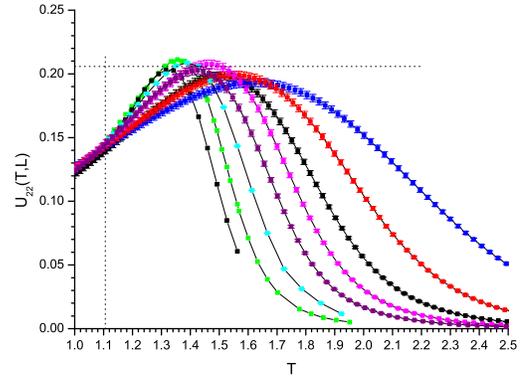}
  \caption{(Color on line) Bimodal 3D ISG. Non-self averaging
    parameter $U_{22}(T,L)$ against temperature $T$. $L=4$, $6$, $8$,
    $10$, $12$, $16$, $20$, $24$ (blue, red, black, pink, purple,
    cyan, green, brown) right to left. $L = 16$, $20$, $24$ from
    \cite{hasenbusch:08}. Vertical line $T_{c}$.}
  \protect\label{fig8}
\end{figure}

When scaled against $\xi(T,L)/L$, in the paramagnetic range
$U_{22}(T,L) =[K_{3}\xi(T,L)/L]^3$ following the Aharony-Harris law,
with $K_{3} = 2.6(1)$, Fig.~\ref{fig7}. The peak locations correspond to
$\xi(T,L)/L \approx 0.35$.

\begin{figure}
  \includegraphics[width=3.0in]{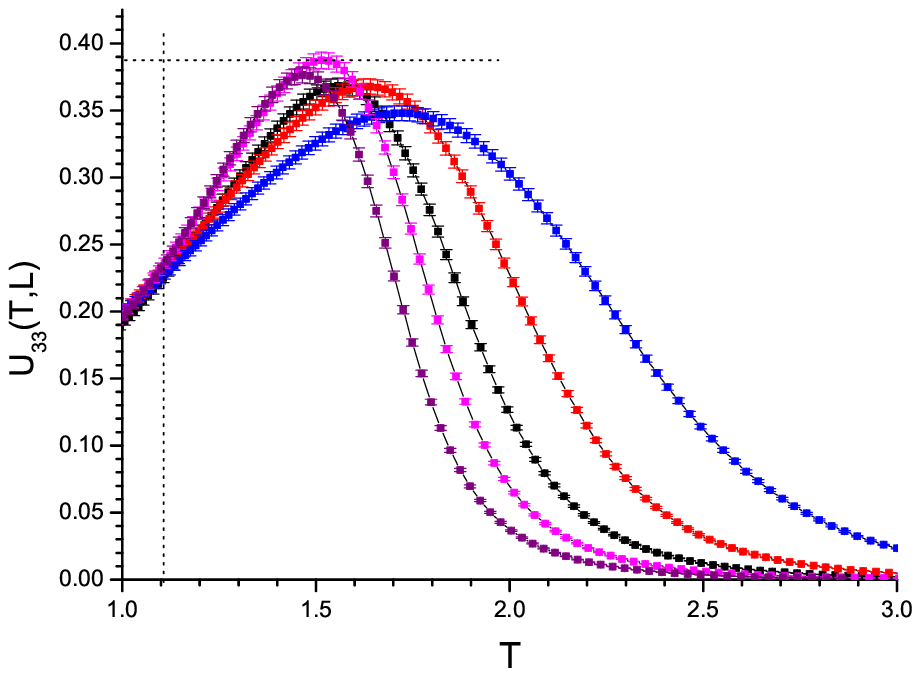}
  \caption{(Color on line) Bimodal 3D ISG. Non-self-averaging
    parameter $U_{33}(T,L)$ against temperature $T$. $L = 4$, $6$,
    $8$, $10$, $12$ (blue, red, black, pink, purple) right to left.
    Vertical line $T_{c}$.} \protect\label{fig9}
\end{figure}

\begin{figure}
  \includegraphics[width=3.0in]{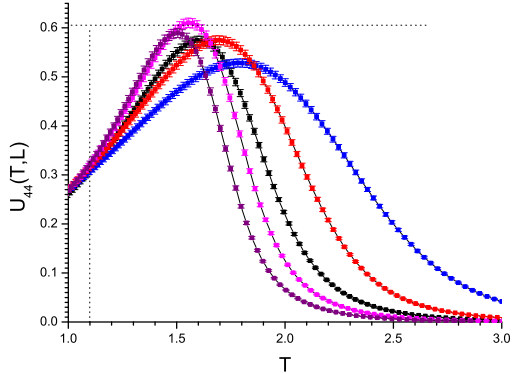}
  \caption{(Color on line) Bimodal 3D ISG. Non-self-averaging
    parameter $U_{44}(T,L)$ against temperature $T$. $L = 4$, $6$,
    $8$, $10$, $12$ (blue, red, black, pink, purple) right to left.
    Vertical line $T_{c}$.} \protect\label{fig10}
\end{figure}

In the large $L$ limit, the $U_{33}$ and $U_{44}$ peak locations are
tending to $T_{c}$, and the peak values extrapolate to $U_{33} \sim
0.39$ and $U_{44} \sim 0.61$, Fig.~\ref{fig9} and \ref{fig10}.

\section{Dimension 4}

\begin{figure}
  \includegraphics[width=3.0in]{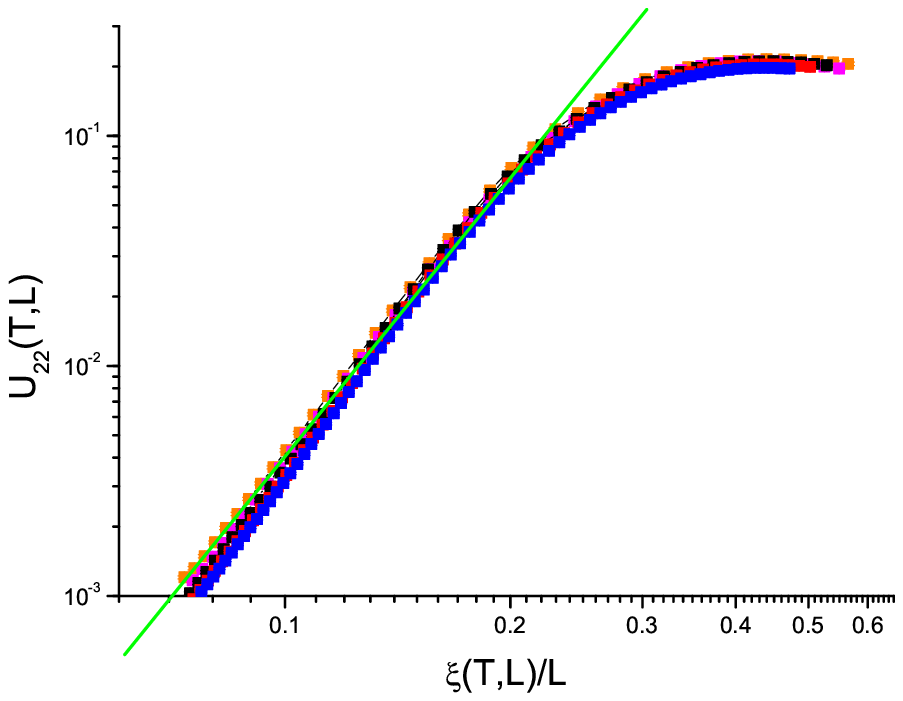}
  \caption{(Color on line) Gaussian 4D ISG. Non-self-averaging
    parameter $U_{22}(T,L)$ against the normalized correlation length
    $\xi/LT,L)/L$.  $L = 4$, $6$, $8$, $10$, $12$ (blue, red, black,
    pink, green).  The straight line has slope
    4.}\protect\label{fig11}
\end{figure}

\begin{figure}
  \includegraphics[width=3.0in]{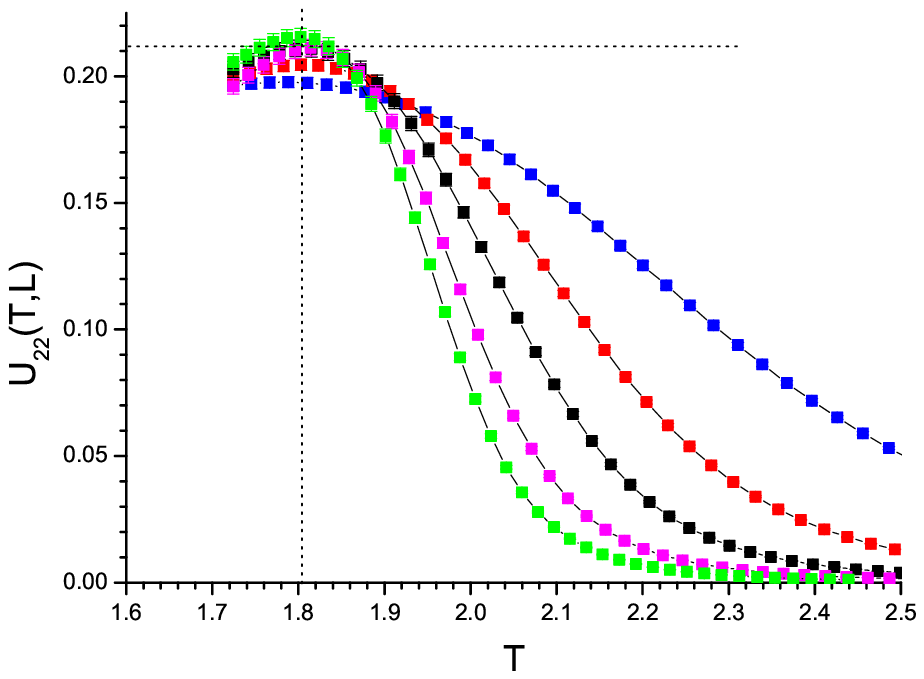}
  \caption{(Color on line) Gaussian 4D ISG. Non-self-averaging
    parameter $U_{22}(T,L)$ against the temperature $T$. $L = 4$, $6$,
    $8$, $10$, $12$ (blue, red, black, pink, green) from right to
    left.  The horizontal line is an extrapolation to criticality at
    $T = T_{c} $ (vertical line).}\protect\label{fig12}
\end{figure}

\begin{figure}
  \includegraphics[width=3.0in]{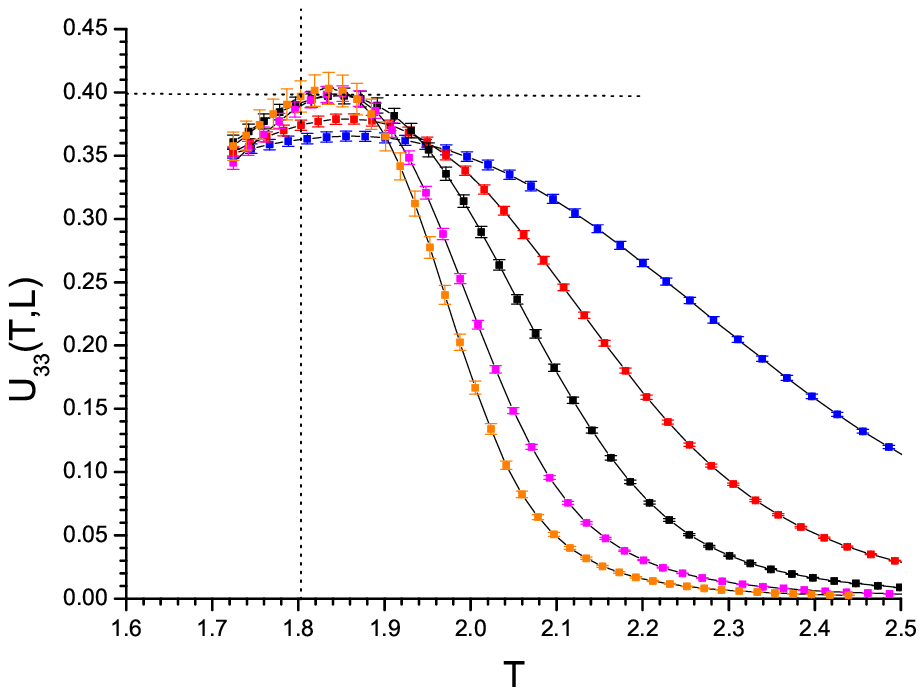}
  \caption{(Color on line) Gaussian 4D ISG. Non-self-averaging
    parameter $U_{33}(T,L)$ against the temperature $T$.$ L = 4$, $6$,
    $8$, $10$, $12$ (blue, red, black, pink, green) from right to
    left.  The horizontal line is an extrapolation to criticality at
    $T = T_c$ (vertical line).} \protect\label{fig13}
\end{figure}

\begin{figure}
  \includegraphics[width=3.0in]{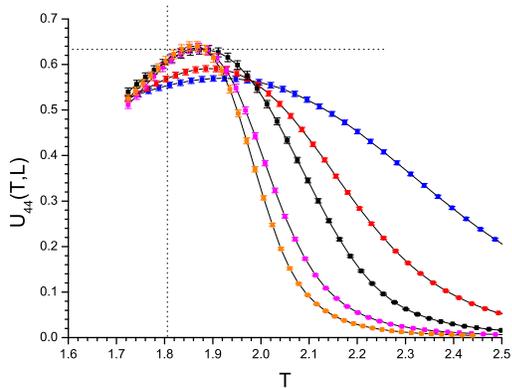}
  \caption{(Color on line) Gaussian 4D ISG. Non-self-averaging
    parameter $U_{44}(T,L)$ against the temperature $T$ . $L = 4$,
    $6$, $8$, $10$, $12$ (blue, red, black, pink, green) from right to
    left.  The horizontal line is an extrapolation to criticality at
    $T = T_c$ (vertical line).} \protect\label{fig14}
\end{figure}

$U_{22}(T,L), U_{33}(T,L)$ and $U_{44}(T,L)$ data for the Gaussian ISG
in dimension four are shown in Figs.~\ref{fig11}, \ref{fig12},
\ref{fig13}, \ref{fig14}. The data correspond to $N=8192$ samples for
each $L$. The critical temperature is $T_c = 1.80(3)$
\cite{katzgraber:06,lundow:15} and the finite size critical value for
the normalized correlation length ratio $[\xi/L]_{c}= 0.440(5)$
\cite{katzgraber:06,lundow:15}. Scaling against the normalized
correlation length Fig.~\ref{fig11}, $U_{22}(T,L)
=(K_{4}\xi(T,L)/L)^4$ again following the Aharony-Harris law, with
$K_{4} = 2.7(1)$ and peaks located at $\xi(T,L)/L=0.43(2)$ so very
close to $\xi(T_{c},L)/L$.
 
Data obtained for the $4$D bimodal ISG (not shown) follow a very
similar pattern with the same peak height.  The locations of the
$U_{nn}(T,L)$ peaks are almost independent of $L$. This was noted for
$U_{22}(T,L)$ in the $4$D Gaussian ISG in Ref.~\cite{palassini:03},
and in Ref.~\cite{jorg:08} for a bond-diluted bimodal model; it
follows from the proximity of the peak $\xi(T,L)/L$ and critical
$\xi(T_{c},L)/L$ values. For the bond-diluted bimodal model, the peak
height is $\approx 0.205$ \cite{jorg:08}.  Because of the
quasi-$L$-independence, the peak location extrapolated to infinite
size provides an estimate for $T_c$ which is limited in precision only
by the statistical uncertainties.

The Gaussian $U_{nn}(T,L)$ peak heights become independent of $L$ to
within the statistical errors after weak finite size effects for small
$L$, Figs.~\ref{fig12}, \ref{fig13}, \ref{fig14}. The peak height
values $U_{22}(T,L)_{\max} = 0.210(5)$, $U_{33}(T,L)_{\max} =
0.400(5)$, $U_{44}(T,L)_{\max} = 0.63(2)$ are the same as those in
dimensions $2$ and $3$ to within the statistical precision. The
stability of the $U_{nn}(T,L)$ peak heights as $L$ is varied turns out
to be a useful empirical criterion for the quality of equilibration.

Alternatively, considering the $U_{nn}(T,L)$ as dimensionless
variables, the intersections of the curves for fixed $L$ should also
give a criterion for estimating $T_{c}$, but the statistical
fluctuations and corrections to scaling affect the intersections much
more drastically than they do the peak location, which means that this
is an imprecise criterion in the $4$D case as noted by \cite{jorg:08}.

\section{Dimension 5}

\begin{figure}
  \includegraphics[width=3.0in]{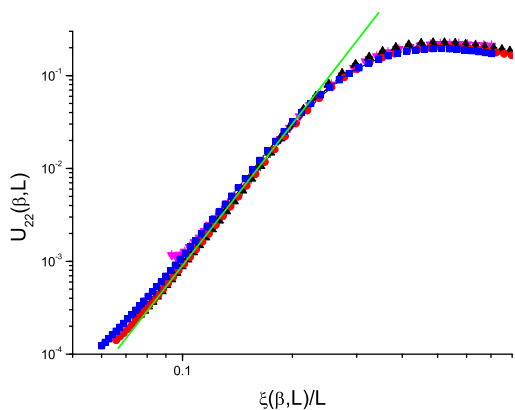}
  \caption{(Color on line) Gaussian 5D ISG. Non-self-averaging
    parameter $U_{22}(T,L)$ against the normalized correlation length
    $\xi(T,L)/L$. $L = 4$, $6$, $8$, $10$ (blue, red, black, pink). The
    straight line has slope 5.} \protect\label{fig15}
\end{figure}

\begin{figure}
  \includegraphics[width=3.0in]{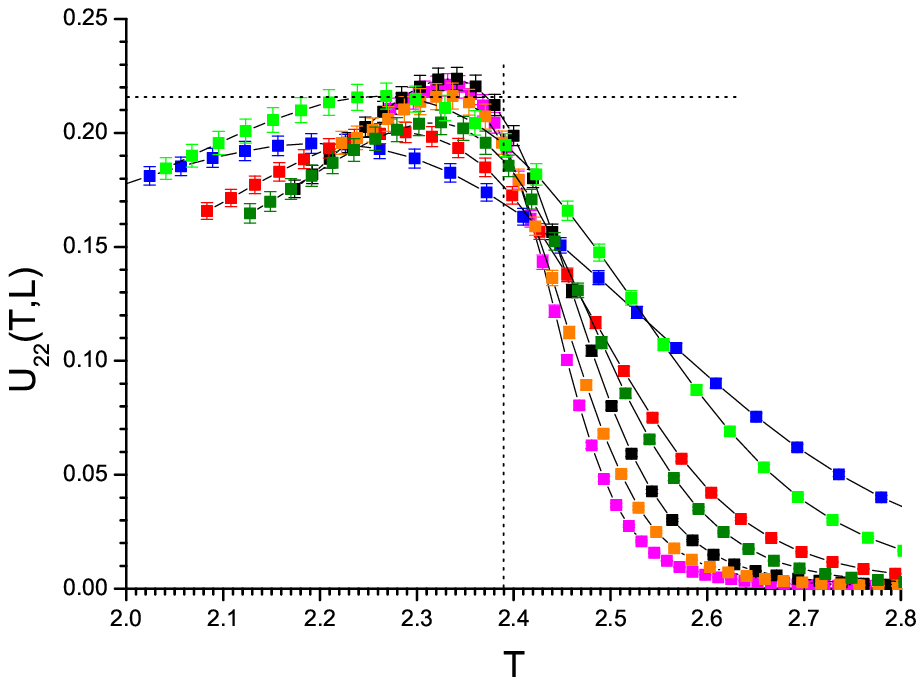}
  \caption{(Color on line) Gaussian 5D ISG. Non-self-averaging
    parameter $U_{22}(T,L)$ against the temperature$ T$. $L = 4$, $5$,
    $6$, $7$, $8$, $9$, $10$ (blue, green, red, olive, black, orange,
    pink) from right to left on the right. The horizontal line is an
    extrapolation to criticality at $T = T_c$. (vertical line)
  }\protect\label{fig16}
\end{figure}

\begin{figure}
  \includegraphics[width=3.0in]{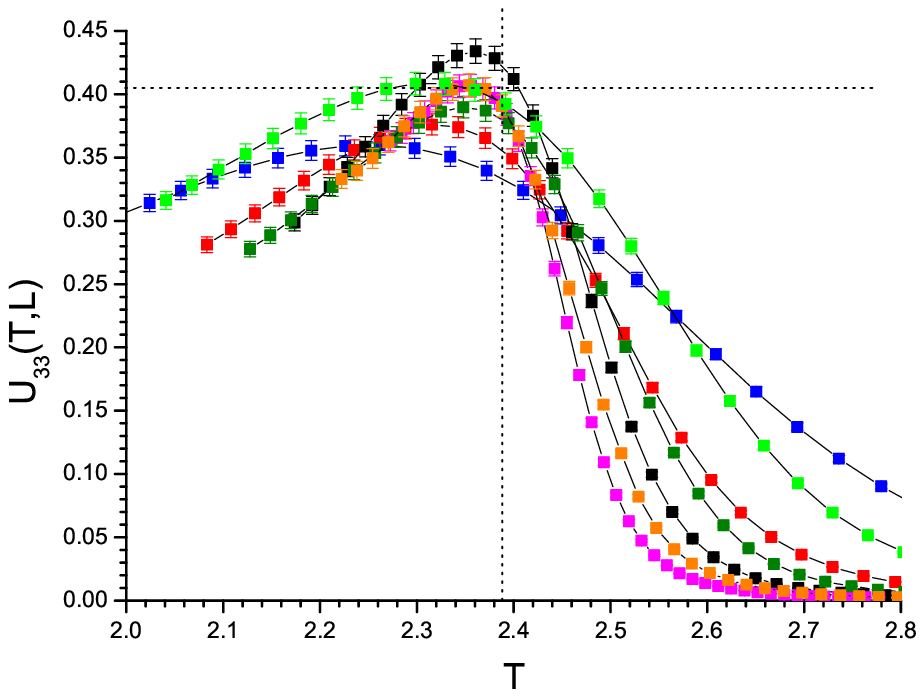}
  \caption{(Color on line) Gaussian 5D ISG. Non-self-averaging
    parameter $U_{33}(T,L)$ against the temperature $T$.$ L = 4$, $5$,
    $6$, $7$, $8$, $9$, $10$ (blue, green, red, olive, black, orange,
    pink) from right to left on the right. The horizontal line is an
    extrapolation to criticality at $T = T_c$. (vertical line)}
  \protect\label{fig17}
\end{figure}

\begin{figure}
  \includegraphics[width=3.0in]{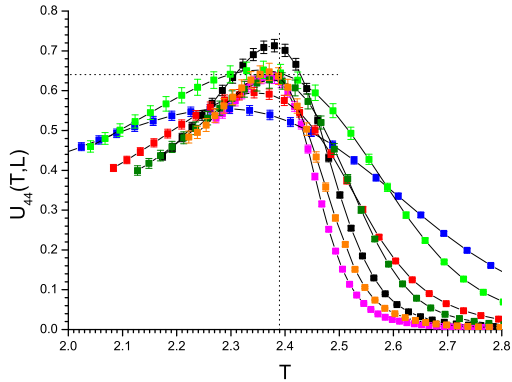}
  \caption{(Color on line) Gaussian 5D ISG. Non-self-averaging
    parameter $U_{44}(T,L)$ against the temperature $T$. $L = 4$, $5$,
    $6$, $7$, $8$, $9$, $10$ (blue, green, red, olive, black, orange,
    pink) from right to left on the right. The horizontal line is an
    extrapolation to criticality at $T = T_c$. (vertical line)}
  \protect\label{fig18}
\end{figure}

$U_{22}(T,L), U_{33}(T,L)$ and $U_{44}(T,L)$ data for the Gaussian ISG
in dimension five are shown in Figs.~\ref{fig15}, \ref{fig16},
\ref{fig17}, \ref{fig18}. The data correspond to $4096$ samples for
each $L$. The critical temperature is $T_c = 2.390(5)$ and the finite
size critical value for the normalized correlation length ratio
$[\xi/L]_{c} \approx 0.45$ \cite{lundow:15b}. We are not aware of
other comparable simulation measurements in this dimension. Data
obtained for the $5$D bimodal ISG (not shown) are very similar.  The
$U_{nn}(T,L)$ peak heights become independent of $L$ to within the
statistical errors after weak finite size effects for small $L$. The
peak height values $U_{22}(T,L)_{\max} = 0.215(5)$,
$U_{33}(T,L)_{\max} = 0.405(5)$, $U_{44}(T,L)_{\max} = 0.64(2)$ are
again practically the same as those in dimensions $2, 3$ and $4$ to
within the statistical precision.

When scaled against the correlation length ratio, in the paramagnetic
range $U_{22}(T,L) = [K_{5}\xi(T,L)/L]^5$ following the Aharony-Harris
law \cite{aharony:96}, with $K_{5} = 2.5(1)$. The peak locations
correspond to $\xi(T,L)/L \approx 0.50 $.  As this ratio is greater
than $[\xi/L]_{c}$ the locations of the $U{nn}(T,L)$ peaks are at
temperatures below $T_{c}$ and the peak temperatures move upwards
towards $T_{c}$ with increasing $L$.  The peak location extrapolated
to infinite size provides an estimate for $T_c$ which is again limited
by the statistical precision but which provides a useful independent
check on the value of the ordering temperature.

\section{Dimension 7}

By this dimension, $N$ the number of spins per sample has become very
large, ($N=823,543$ for $L=7$), which imposes practical limits on the
sizes and numbers of samples in the simulations. The simulations were
carried out for $L=3$ to $7$ with $512$ samples at each $L$.

\begin{figure}
  \includegraphics[width=3.0in]{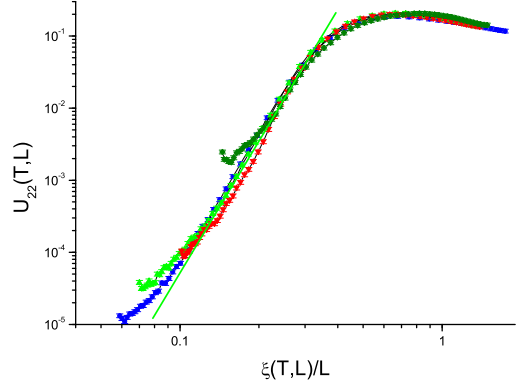}
  \caption{(Color on line) Bimodal 7D ISG. Non-self-averaging
    parameter $U_{22}(T,L)$ against the normalized correlation length
    $\xi(T,L)/L$. $L = 4$, $5$, $6$, $7$ (blue, green, red, olive).
    The straight line has slope 6.}
  \protect\label{fig19}
\end{figure}

The dimension $7$ bimodal ISG has an ordering temperature
$T_{c}=3.39(1)$ estimated using the standard Binder cumulant crossing point
technique \cite{lundow:15b} in agreement with the high temperature
series expansion (HTSE) estimates $T_{c} = 3.37(2)$ \cite{singh:86}
and $T_{c}= 3.384(15)$\cite{daboul:04}. (Curiously the HTSE value
given in Ref.~\cite{klein:91} corresponds to $T_{c}= 3.459$. We
suspect a typographical error).  As this dimension is above the upper
critical dimension $d=6$, the critical exponents $\gamma=1$ and $\nu =
1/2$ are known exactly. In this case in the paramagnetic regime
$U_{22}(T,L) = (K_{7} \xi(T,L)/L)^{6}$, with an exponent which appears
to be $\approx 6$ rather than $7$, Fig.~\ref{fig19}. This could arise
from the breakdown of the relations between scaling exponents above
the ucd. Because of the limited number of samples and the small values
of $L$ at this dimension, this estimate is not very precise.

\begin{figure}
  \includegraphics[width=3.0in]{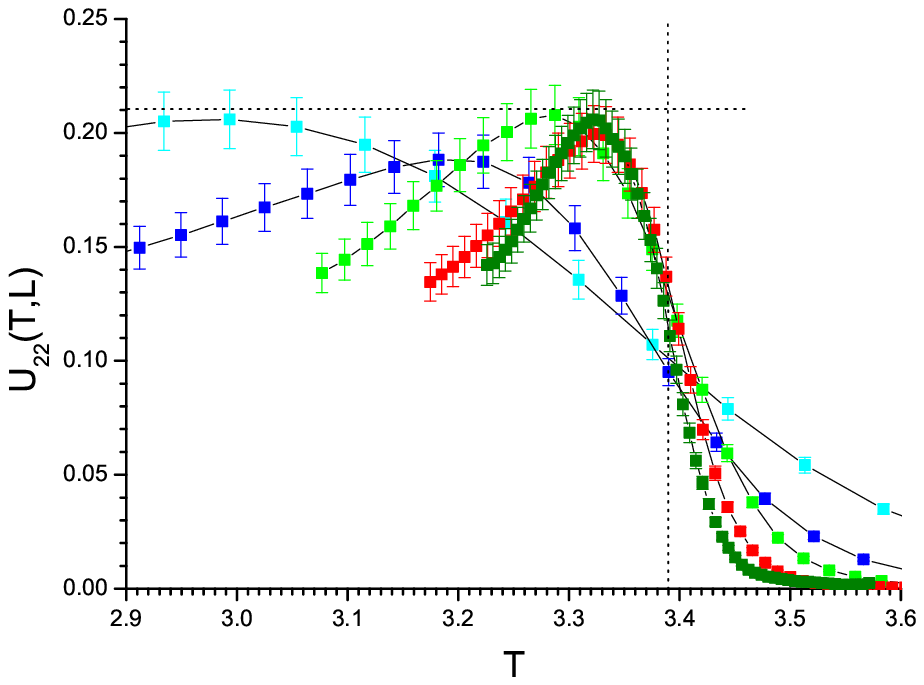}
  \caption{(Color on line) Bimodal 7D ISG. Non-self averaging
    parameter $U_{22}(T,L)$ against temperature $T$. $L = 3$, $4$,
    $5$, $6$, $7$ (cyan, blue, green, red, olive) right to left.
    Vertical line $T_{c}$.} \protect\label{fig20}
\end{figure}

In the plot of $U_{22}(T,L)$ against $T$, Fig.~\ref{fig20}, the
$L$-independent critical finite size crossing point value is
$U_{22}(T_{c}) \approx 0.15$, and the $[U_{22}(T,L)]_{\max}$ peak
heights are independent of $L$ and equal to $\approx 0.21$ to within
the statistics, as for the other dimensions. The maxima locations move
towards $T_c$ from within the ordered regime. This behavior is very
similar to that observed in the mean field ISG SK model
\cite{hukushima:00,picco:01}.

\begin{figure}
  \includegraphics[width=3.0in]{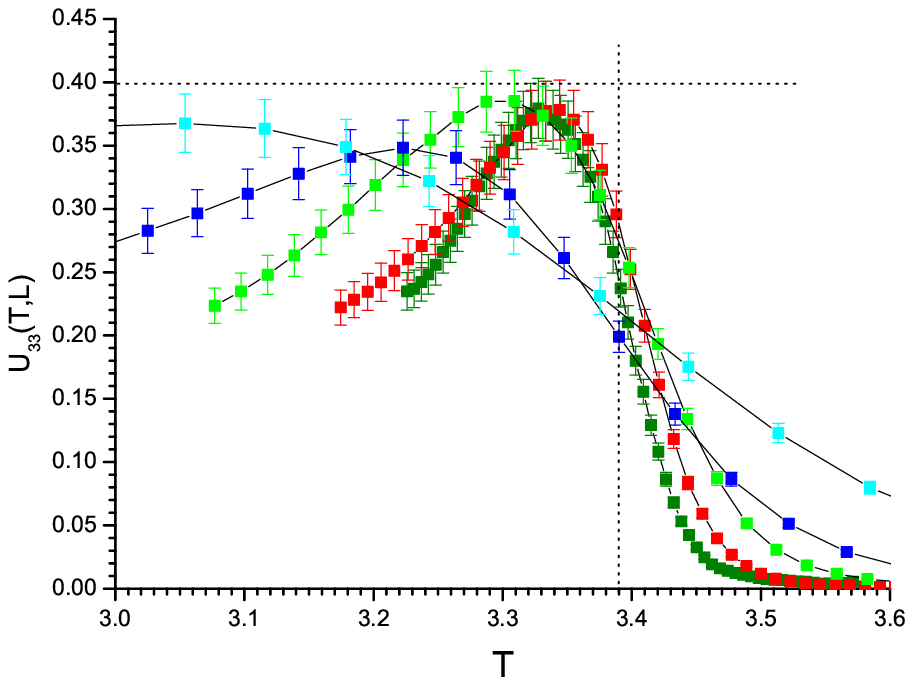}
  \caption{(Color on line) Bimodal 7D ISG. Non-self averaging
    parameter $U_{33}(T,L)$ against temperature $T$. $L = 3$, $4$,
    $5$, $6$, $7$ (cyan, blue, green, red, olive) right to left.
    Vertical line $T_{c}$.} \protect\label{fig21}
\end{figure}

\begin{figure}
  \includegraphics[width=3.0in]{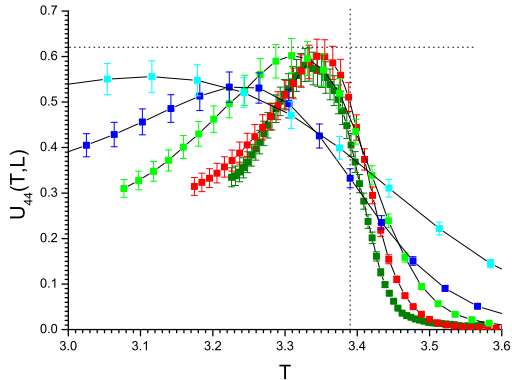}
  \caption{(Color on line) Bimodal 7D ISG. Non-self averaging
    parameter $U_{44}(T,L)$ against temperature $T$. $L = 3$, $4$,
    $5$, $6$, $7$ (cyan, blue, green, red, olive) right to left.
    Vertical line $T_{c}$.} \protect\label{fig22}
\end{figure}

The higher order $U_{33}(T,L)$ and $U_{44}(T,L)$, Fig.~\ref{fig21} and
Fig.~\ref{fig22}, follow much the same pattern, with peak maxima of
$0.39(1)$ and $0.61(2)$ respectively, again equal to the values for
the other dimensions to within the statistics.

\section{The Gauge Glass}

The Gauge glass (GG) is a vector spin glass which does not support
chiral ordering.  The GG in dimension $3$ has a critical temperature
$T_{c}=0.47(1)$ \cite{olson:00,katzgraber:04,alba:11}.  The
non-self-averaging parameter $U_{22}(L,T)$ scales with $\xi(L,T)/L$
\cite{alba:11} and shows a maximum peak height independent of $L$ and
a peak position $T_{\max}(L)$ near $\xi(L,T)/L =0.35$. The
paramagnetic regime data \cite{alba:11} appear by inspection to be
compatible with the Aharony-Harris rule $U_{22}(L,T) \sim
[\xi(L,T)/L]^{3}$ although the published data are not presented in
this way.  As the critical correlation length ratio is
$[\xi(L,T)/L]_{c} =0.54$ \cite{alba:11}, the $U_{22}(L,T)$ peak
temperature location moves downwards with $L$ and tends towards
$T_{c}$. The $3$D vector spin glass GG $U_{22}(T,L)$ thus follows
basically the same rules as followed by $U_{22}(T,L)$ in the Ising
spin glass in $3$D, except that the GG peak maximum is $\approx 0.10$
instead of $0.205$. Data on GGs in dimensions $2,3$ and $4$ from
measurements which were not designed to estimate the
non-self-averaging parameter \cite{katzgraber:04} are consistent with
$U_{22}(T,L)$ peak values near $0.10$ in each dimension. We can
speculate that this family of spin glass models also has its
characteristic dimension-independent value of the non-self-averaging
peak height.

\section{Heisenberg and XY spin glasses}

Numerical measurements on Heisenberg spin glasses (HSGs) are of
particular importance because the canonical experimental spin glass
dilute alloys ({\bf Au}Fe, {\bf Cu}Mn, {\bf Ag}Mn) are all Heisenberg
systems, so it should be possible to understand the ordering mechanism
in "real life" spin glasses on the basis of numerical data on
Heisenberg models. We have no new data to report on these models but
it is of interest to consider published non-self-averaging data in the
light of the ISG results.

Both Heisenberg and XY spin glasses can support chiral glass order as
well as spin glass order, and for many years there have been two
conflicting interpretations of the numerical data on the ordering
transitions in these models in dimension $3$. According to the first
interpretation, the ordering is spin-spin interaction driven;
basically the ordering process is much the same as in ISGs, and the
chiral order follows on as a geometrically necessary consequence of
the onset of spin order, without the chiral interactions playing any
significant role in the spin glass transition
\cite{lee:03,campos:06,lee:07,pixley:08,fernandez:09}. The alternative
interpretation is that the driving role in $3$D HSG or XYSG ordering
is played by the chirality, so that there is first a chiral order
onset followed at a lower temperature by spin ordering transition
\cite{kawamura:10,hukushima:05,viet:09,obuchi:13}. (Similar
disagreements concerning fully frustrated $2$D XY models were resolved
definitively in favor of a distinct chiral order transition just above
a spin order transition \cite{hasenbusch:05,okumura:11}). The
arguments of both schools to support their respective interpretations
in the $3$D HSG and XYSG models have been essentially based on
analyses of the data for the crossing points of the dimensionless
normalized spin and chiral ($s$ and $c$) correlation lengths
$\xi_{s}(T,L)/L$ and $\xi_{c}(T,L)/L$. The numerical simulations in
the spin glasses are even more demanding than in the fully frustrated
models, and because of intrinsic finite size corrections and the need
to reach strict equilibration at each $L$, extrapolations to infinite
$L$ in order to estimate the ThL crossing point locations are
delicate. As simulations were extended to larger sizes in successive
Gaussian HSG and XYSG measurements interpreted on the spin-driven
ordering scenario, the joint spin/chiral crossover temperature was
estimated to be $T_{c}(HSG) \approx 0.160$ \cite{lee:03}, $T_{c}(HSG)
\approx 0.145$ with a KTB-like critical line \cite{campos:06},
marginal but very similar spin and chiral behavior (XYSG and
HSG)\cite{lee:07,pixley:08}, and most recently $T_{c}(HSG) \approx
0.120$ \cite{fernandez:09}. No non-self-averaging results were
reported. From detailed $3$D bimodal and Gaussian HSG and $3$D
Gaussian XYSG measurements, the two separate transition temperatures
on the chiral-driven ordering scenario are estimated to be (bimodal
HSG) \cite{hukushima:05}, $T_c(c)=0.194(5)$ and $T_c(s)\le 0.15$,
(Gaussian HSG) $T_{c}(c) = 0.143(3)$ and $T_{c}(s) =
0.125(+0.006/-0.012)$ \cite{viet:09}, and (XYSG) $T_{c}(c) = 0.308(5)$
and $T_{c}(s) = 0.274(3)$ \cite{obuchi:13}. Non-self-averaging data
were shown in each case.

In the light of the ISG results reported above, it would appear that
in Heisenberg and XY spin glasses non-self-averaging could provide an
independent primary numerical criterion for spin and/or chiral
ordering much less sensitive to finite size effects and to strict
equilibration (as already suggested in Ref.~\cite{hukushima:05}). On
the first (spin-driven ordering) scenario one would expect the spin
non-self-averaging parameter $U_{22s}(T,L)$ to follow much the same
rules as for the ISG or the GG chiral-free vector spin glass cases
discussed above, with a peak location moving towards an upper spin
ordering temperature $T_{c}(s)$ as $L$ increases, and a regular
behavior reflecting $U_{22s}(T,L) \sim (\xi_{s}(T,L)/L)^{3}$ in the
paramagnetic regime above $T_{c}(s)$. On this interpretation the
chiral $U_{22c}(T,L)$ would be weaker than the $U_{22s}(T,L)$; if a
$U_{22c}(T,L)$ peak exists, it would be located at a temperature below
or possibly at the ThL $U_{22s}(T,L)$ peak.

On the second (chiral-driven order) scenario, it would be the chiral
$U_{22c}(T,L)$ which would show a peak first, with a peak location
tending towards the (upper) chiral ordering temperature $T_{c}(c)$ as
$L$ increases. In the paramagnetic regime one would expect a regular
behavior of the chiral non-self-ordering $U_{22c}(T,L)$ with
increasing $L$, governed by $U_{22c}(T,L) \sim
(\xi_{c}(T,L)/L)^{3}$. On this scenario the spin $U_{22s}(T,L)$ would
then show a peak location somewhere below the chiral $U_{22c}(T,L)$
peak, with a location tending towards an ordering temperature
$T_{c}(s)$ below $T_{c}(c)$, together with a paramagnetic regime
$U_{22s}(T,L)$ behavior behaving irregularly at least at small $L$
because the paramagnetic spin ordering is perturbed by the dominant
onset of chiral order.

Very informative non-self-averaging data have been published on the
$3$D HSG with bimodal interactions \cite{hukushima:05}, on the $3$D
HSG with Gaussian interactions \cite{viet:09}, and on the $3$D
Gaussian XYSG \cite{obuchi:13}. In each case the pattern is the same :

- there is a strong $U_{22c}(T,L)$ peak at an almost $L$-independent
temperature $T \approx 0.19$, $T \approx 0.145$, $T \approx 0.31$
respectively, so in each case close to the $T_{c}(c)$ value estimated
independently from other criteria
\cite{hukushima:05,viet:09,obuchi:13}. In the paramagnetic regime
there is a regular narrowing in temperature of the $U_{22c}(T,L)$ peak
with increasing $L$ which appears compatible with the Aharony-Harris
law $U_{22c}(T,L) \sim (\xi_{c}(T,L)/L)^{3}$ though the data are not
presented in this form.

 - in each case, the spin $U_{22s}(T,L)$ peak is either not visible
 (HSGs) or is marginal (XYSG) down to the lowest temperature at which
 non-self-averaging measurements were made, $T \approx 0.145$,$T
 \approx 0.11$ to $0.13$ depending on $L$, and $T \approx 0.24$ to
 $0.275$ depending on $L$ in the three cases. Over the whole
 temperature range $U_{22s}(T,L)$ is always weaker than
 $U_{22c}(T,L)$, and $U_{22s}(T,L)$ has irregular behavior as a
 function of $L$ in the paramagnetic temperature regime at and above
 the $U_{22c}(T,L)$ peak.

Thus the non-self-averaging data $U_{22c}(T,L)$ and $U_{22s}(T,L)$ in
the three models \cite{hukushima:05,viet:09,obuchi:13} can be seen by
inspection to be clearly incompatible with the behavior expected on
the spin-driven ordering scenarios
\cite{lee:03,campos:06,lee:07,pixley:08,fernandez:09}, and fully
compatible with the chiral-driven ordering
interpretation\cite{kawamura:10,hukushima:05,viet:09,obuchi:13}.

\section{Conclusion}

The non-self-averaging data on ISGs in all dimensions show a
remarkable regularity. In each dimension there is a peak as a function
of temperature in the standard non-self-averaging parameter
$U_{22}(T,L)$ and in the higher order parameters $U_{33}(T,L)$ and
$U_{44}(T,L)$ whose values are $L$-independent after weak small size
effects; the peak values $U_{22}(T,L)_{\max} \approx 0.21$,
$U_{33}(T,L)_{\max} \approx 0.40$, $U_{44}(T,L)_{\max} \approx 0.62$
are independent of dimension to within the statistics for dimensions
$2, 3, 4, 5$ and $7$. In the paramagnetic regime above the peak the
Aharony-Harris renormalization group law \cite{aharony:96}
$U_{nn}(T,L) = (K_{d}\xi(T,L)/L)^d$ is obeyed, with $K_{d}(U_{22})
\approx 2.6$ for all dimensions. Both of these empirical observations
can be classed as \lq\lq hyperuniversal behavior\rq\rq{}.

Published \cite{alba:11} and unpublished \cite{katzgraber:04} data on
the Gauge Glass, a vector spin glass which does not support chirality,
suggest that non-self-averaging rules analogous to those that hold in
the ISGs appear to apply but with a different characteristic peak
height $U_{22}(T,L)_{\max} \approx 0.10$.

XY and Heisenberg spin glasses can support chiral ordering as well as
spin ordering. In the light of the non-self-averaging behavior
reported above for the ISG models, it is clear that the published spin
and chiral non-self-averaging data
\cite{hukushima:05,viet:09,obuchi:13} in $3$D Heisenberg and $XY$
models are incompatible with a spin-driven ordering scenario
\cite{lee:03,campos:06,lee:07,pixley:08,fernandez:09} but strongly
support the alternative conclusion that the spin glass ordering in
these models is chiral-driven rather than spin-driven, on the Kawamura
scenario \cite{kawamura:10}. An important implication is that order in
the canonical experimental Heisenberg spin glasses is also chirality
driven.

\begin{acknowledgments}
  We would like to thank H. Kawamura for helpful comments.  The
  computations were performed on resources provided by the Swedish
  National Infrastructure for Computing (SNIC) at the High Performance
  Computing Center North (HPC2N) and Chalmers Centre for Computational
  Science and Engineering (C3SE).
\end{acknowledgments}

\end{document}